\documentclass[prd,twocolumn,floatfix,reprint,amsmath,nofootinbib,superscriptaddress,amssymb,aps,preprintnumbers,floatfix]{revtex4-1}
 \usepackage{amsmath,txfonts,longtable,booktabs,overpic,amssymb,bm,bbm,multirow,float,graphicx,color,dcolumn,subfigure,hyperref,tikz,fancyhdr}
 \usepackage{soul}
\definecolor{blue}{RGB}{45,48,146}
  \hypersetup{colorlinks,citecolor=blue,anchorcolor=red,menucolor=red, linkcolor=red,filecolor=red,runcolor=red,urlcolor=blue,frenchlinks=true}

\begin{document}

\title{Spin effect in vacuum pair production under two-color rotating electric fields}
\author{Zhao-Yuan Chen}
\affiliation{Xinjiang Key Laboratory of Solid State Physics and Devices, School of Physics Science and Technology, \href{https://ror.org/059gw8r13}{Xinjiang University}, Urumqi 830017, China}
\author{Orkash Amat}
\affiliation{ School of Astronomy and Space Science, \href{https://ror.org/01rxvg760}{Nanjing University}, Nanjing 210023, China}
\author{Jin-hui Bai}
\affiliation{Xinjiang Key Laboratory of Solid State Physics and Devices, School of Physics Science and Technology, \href{https://ror.org/059gw8r13}{Xinjiang University}, Urumqi 830017, China}
\author{Mamat Ali Bake}
\email[Electronic mail:]{mabake@xju.edu.cn}
\affiliation{Xinjiang Key Laboratory of Solid State Physics and Devices, School of Physics Science and Technology, \href{https://ror.org/059gw8r13}{Xinjiang University}, Urumqi 830017, China}


\begin{abstract}
We investigated the spin effect on the vacuum pair production by Dirac-Heisenberg-Wigner (DHW) formalism under two-color counter-rotating electric fields. We primarily studied the combined effects of the field asymmetry, time delay, and frequency chirp on the particle momentum spectrum with and without considering the spin effect.
We have observed that the vacuum pair production process demonstrates spin dependence even in a pure electric field and is sensitive to variations in the field parameters.
The results indicate that the spin-dependent momentum spectrum exhibited distinct outcomes for various asymmetric fields with different chirp values and time delay.
For an extended asymmetric field with large chirp and time delay, the particle number density can be increased by more than six orders of magnitude.
The spin-up and spin-down particles are approximately comparable for a symmetric field with a small-frequency chirp and are dominated by the spin-up particles for a larger chirp.
However, in the case of an asymmetric field, the increase in field asymmetry and the chirp parameter lead to a reversal of the spin asymmetry degree. For a shortened asymmetric electric field with a large-frequency chirp, the number of spin-up particles increases, leading to a spin asymmetry degree of $98.62\%$. Conversely, in an extended asymmetric field, the number of spin-down particles increases significantly, which corresponds to a spin asymmetry degree of $99.94\%$.
\end{abstract}
\maketitle

\section{Introduction}
Quantum electrodynamics (QED) predicts matter creation from light in a quantum vacuum, such as electron-positron pair production under strong electromagnetic fields. This prediction was first proposed by Sauter in 1931 \cite{Sauter:1931zz} and later extended by Schwinger \cite{Schwinger:1951nm}, hence the phenomenon is also known as the Sauter-Schwinger effect. However, due to the extremely high electric field strength $E_{cr}\approx 1.3\times 10^{18}~\textrm{V}/\textrm{m}$ (so-called Schwinger critical field strength, the corresponding laser intensity is $I_{cr}\sim 4.3 \times 10^{29}~\text{W/cm}^{2}$) required to produce electron--positron pairs, researchers have not been able to observe this phenomenon experimentally. The Chirped Pulse Amplification (CPA) technique proposed by D. Strickland and G. Mourou for the first time, which results in a quick increase in power by 1000 times \cite{Strickland:1985gxr}, has continuously improved in recent years, and the intensity of high-power lasers is expected to approach $I\sim 10^{26}~\text{W/cm}^{2}$ in the near future. This allows researchers to observe the nonperturbative Schwinger effect in the laboratory \cite{eli-beams,xcels}.

As researchers study matter-antimatter production in strong electric fields with varying frequencies and strengths, they find that electron-positron pairs are produced via two distinct mechanisms distinguished by the Keldysh adiabaticity parameter $\gamma = {m_e\omega} / {eE_0}$ (where $m_e$, $e$, $\omega$, and $E_0$ represent the electron mass, charge, frequency, and external electric field strength, respectively). When the external electric field is a low-frequency strong field $\gamma \ll 1$, electron--positron pairs are primarily produced through the tunneling process. By contrast, for a high-frequency weak field $\gamma \gg 1$, the electron--positron pairs are mainly created via the multiphoton absorption mechanism. Multiphoton absorption was observed in SLAC experiments by the collision of a high-energy electron beam with an intense laser pulse \cite{SLAC1,SLAC2}. However, the tunneling process has not yet been experimentally explored because the laser intensity achievable in modern laboratories has not yet reached the Schwinger critical field strength. Recent studies have shown that when two fields are combined, the yield of electron--positron pairs is greatly enhanced by the dynamically assisted Schwinger mechanism \cite{Schutzhold:2008pz,Taya:2020,CK:2021}, which is an efficient approach for enhancing the pair production rate in the subcritical Schwinger limit. Many previous studies have verified that, in subcritical field situations, where each mechanism is separately suppressed, their combined effect significantly enhances the pair production rate through dynamical assistance, which is equivalent to reducing the required field intensity \cite{Aleksandrov2018,IbrahimSitiwaldi2018}. Therefore, the effects of various field combinations with different field frequencies and intensities on vacuum-pair production have begun to attract attention. In 2010, researchers used the WKB scattering approach to study the impact of chirp parameters and carrier phase on the momentum spectrum of the produced particles \cite{Dumlu2010}. Subsequently, an increasing number of studies have been conducted, including the modulation of the pulse shape, variation in the number of laser pulses, and change in frequency (frequency chirp) \cite{chirped1,chirped2,chirped3,chirped4,chirped5,chirped6,chirped7}.

Although enhancement of the creation rate is the main concern in pair production studies, the momentum spectra of pairs could be helpful for understanding the dynamics of the problem from both theoretical and experimental points of view. Therefore, the exploration of external field configurations in the pair production mechanism has been of great interest in recent years, and some interesting phenomena in the particle momentum spectrum have been discovered. By combining the two rotating electric fields, the momentum spectrum of the generated particles exhibited a vortex structure \cite{vortices2}. The appearance of the vortex structure is a direct result of the combination of two counter-rotating fields. However, the shape of the vortex structure is influenced by many different factors, such as the ellipticity of the electric field, frequency, number of periods, phase, and time delay \cite{vortices1,vortices2,vortices3,vortices4}.

In recent research, the spin effect in electrodynamics has gradually come into focus and been widely investigated in laser-electron/plasma interactions and vacuum pair production processes. In 2015, Strobel et al. studied the vacuum pair production under a two-component time-varying electric field, revealing that for rotating electric fields, the pair creation rate is dominated by particles with a specific spin, which depends on the sense of rotation as well as the pulse length and frequency within a certain range \cite{Strobel2015}. Subsequently, Blinne et al. compared the vacuum pair production rates and momentum spectra obtained from numerical methods based on the real-time DHW formalism with those obtained from a semiclassical approximation based on scattering matrix analysis \cite{Blinne2016}. These results indicate that the production of pairs with different spin states is not equal in a constant rotating field. Kohlf\"urst studied pair production in rotating electric fields under the multi-photon mechanism and directly obtained the spin-dependent particle production amplitude from the exact numerical solution based on quantum dynamics theory \cite{CKspin2019}. Recently, Hu et al. found that even in a single rotating field, a momentum helix can be induced by the particle spin effect \cite{hulinaspin}. Additionally, significant progress has been made in the analysis of the helicity, chirality, and spin effects in the nonlinear Compton scattering (NCS) and nonlinear Breit-Wheeler (NBW) pair production processes \cite{spin1,spin2,spin3,spin4,spin5}. The momentum distributions of the particles with different chirality and spins represent observable signatures that should facilitate experimental efforts concerning practical investigations of the vacuum pair production and other QED processes \cite{Taya:2020,spin11,spin12,spin13,spin14,spin15,spin16,spin6}.

In previous studies on the vacuum pair production process, the standing wave approximation has predominantly been utilized to neglect the magnetic component of the electromagnetic field. Consequently, spin dynamics within the vacuum pair production process received limited attention as it was commonly assumed that pure electric fields do not couple with spin degrees of freedom. However, recent research suggests that the spin effect in the vacuum pair production process does not require the existence of a magnetic field in the observer frame, indicating that it is purely an electric effect \cite{spin7,spin8,Review}. Some studies have examined the individual impacts of frequency chirp, field asymmetry, and time delay on production efficiency and momentum spectrum with or without spin effects \cite{chirped1,chirped2,chirped3,chirped4,chirped5,chirped6,chirped7,vortices1,
vortices2,vortices3,vortices4,hulinaspin}. However, the combined effects of these parameters and whether optimal combinations exist to enhance pair creation remain unclear. Moreover, to our knowledge, the influence of frequency chirp and field asymmetry on the spin effect in vacuum pair production within the DHW formalism has not been explored.
Experimentally, future investigations into vacuum pair production may potentially be realized using visible light or X-ray lasers \cite{RINGWALD2001107}. To further enhance the laser intensity, chirped pulse amplification techniques could be utilized \cite{Strickland:1985gxr}. Achieving a perfectly symmetric field poses significant experimental challenges. In this context, the combined effects of field asymmetry and frequency chirping are expected to substantially influence the outcomes. Moreover, these parameters will play a critical role in shaping the momentum spectrum and polarization characteristics of the produced electron-positron pairs. Therefore, studying the spin-dependent momentum spectrum in pair production under asymmetric and chirped fields is both theoretically justified and experimentally relevant.

In this study, inspired by the aforementioned investigations, we investigated the particle momentum spectra and spin effect in the vacuum pair production process using two counter-rotating circularly polarized electric fields. We mainly considered the effects of field asymmetry, the time delay between the two fields, and frequency chirp on the particle momentum spectrum, as well as the spin-dependent momentum distribution. For a symmetric electric field without a chirp, the changes in the momentum spectrum, number density, and, more importantly, the spin effect are negligible. However, these effects are considerably enhanced in an asymmetric electric fields with a large chirp and time delay. We found that the particle momentum distribution with the spin effect was highly sensitive not only to the time delay as discussed in Ref. \cite{hulinaspin}, but also to other field parameters, such as the field asymmetry and frequency chirp.

The rest part of this paper is organized as follows. The background-field model is presented in Sec. \ref{field}. The general method of the Dirac-Heisenberg-Wigner (DHW) formalism is briefly introduced in Sec. \ref{DHWformalism}. The numerical calculation results of the DHW equations are provided and the momentum spectra with/without spin effects are discussed in Sec. \ref{results}. Finally, a brief discussion and conclusions are given in Sec. \ref{conclusion}.

\section{ Model of background field}\label{field}
We used the DHW method to investigate the momentum spectra of electron--positron pairs produced by two counter-rotating asymmetric electric fields with time delays under various chirp asymmetry conditions. Our two-color counter-rotating electric field is composed by two spatially homogenous and time-varying rotating electric fields with a little different strength and frequency between them. For the single rotating electric field, the form is given as \cite{chirped3,chirped7,vortices2,vortices3}
\begin{equation}\label{fieldmodel}
\begin{aligned}
   \mathbf{E}_1(t)={{E}_{1}}\left[\exp \left( -\frac{{{t}^{2}}}{2{{\tau }_{11}}^{2}} \right)H(-t)+\exp \left( -\frac{{{t}^{2}}}{2{{\tau }_{12}}^{2}} \right)H(t)\right]
   \\
   \times\left[\begin{matrix}
   \cos \left( {{\omega }_{1}}t+bt^2 \right)  \\
   {{\delta }_{1}}\sin \left( {{\omega }_{1}}t+bt^2 \right)  \\
   0
\end{matrix} \right],
\end{aligned}
\end{equation}
\begin{equation}\label{fieldmode2}
\begin{aligned}
  \mathbf{E}_2(t)={{E}_{2}}\left[\exp \left( -\frac{{{(t-{{T}_{d}})}^{2}}}{2{{\tau }_{21}}^{2}} \right)H(-t)+\exp \left( -\frac{{{(t-{{T}_{d}})}^{2}}}{2{{\tau }_{22}}^{2}} \right)H(t)\right]
  \\
  \times\left[ \begin{matrix}
   \cos ({{\omega }_{2}}(t-{{T}_{d}})+b{{(t-{{T}_{d}})}^{2}})  \\
   {{\delta }_{2}}\sin ({{\omega }_{2}}(t-{{T}_{d}})+b{{(t-{{T}_{d}})}^{2}})  \\
   0  \\
\end{matrix} \right].  \\
\end{aligned}
\end{equation}
And the combined electric field is given by
\begin{equation}\label{fieldmode3}
  \mathbf{E}(t)=\mathbf{E}_1(t)+\mathbf{E}_2(t),
\end{equation}
where $E_{1,2}=E_{01,02}/\sqrt{1+\delta_{1,2}^2}$ represents the strength of the electric fields; $H(t)$ is the Heaviside step function; $\tau_1=\tau_{11}+\tau_{12}$ and $\tau_2=\tau_{21}+\tau_{22}$ are the durations of the electric fields in (\ref{fieldmodel}) and (\ref{fieldmode2}), $\tau_{11,21} = N\pi/\omega_{1,2}$ ($N=2$ is cycle number), and $\tau_{12,22} =k\tau_{11,21}$, where $k$ is the asymmetry ratio parameter; the full duration of the combined field (\ref{fieldmode3}) is $\tau=\tau_1+\tau_2$, $T_d = G\tau$ represents the time delay parameter between two counter-rotating electric fields, and $G$ is a dimensionless quantity; $\omega_{1,2}$ corresponds to the oscillation frequency of the electric fields and $b$ is the chirp parameter, the effective frequency of the electric field was $\omega_{eff}=\omega+bt$; $\lvert\delta_{1,2}\rvert = 1$ represents circular polarization (where we define $\delta = \pm1$ as a right-hand or left-hand circularly polarized field), and we choose $\delta_1 = 1$ and $\delta_2 = -1$ in our calculations \cite{vortices2}. In this study, we set the electric field parameter to $E_{01}=0.1\sqrt{2}E_{cr}$, $E_{02}=0.07\sqrt{2}E_{cr}$, $\omega_1=0.44m$, $\omega_2=0.55m$. Note that the single rotating electric field in Eqs. (\ref{fieldmodel}) and (\ref{fieldmode2}) is obtained from standing-wave field of two counter-propagating short laser pulses with equal wavelength, intensity, and polarization. Then the magnetic fields of the two laser pulses can cancel each other in the standing-wave, and the electric field is approximately spatially homogeneous throughout the entire interaction region. Therefore, the field (\ref{fieldmode3}) can be obtained by combining a rotating electric field (\ref{fieldmodel}) with another time-delayed one (\ref{fieldmode2}).

We used natural units throughout this paper, where $\hbar = c = 1 $, and all physical quantities were expressed in terms of the electron mass $m$. For example, the units of the field frequency and momentum are $m$, and the time scale of the electric field is $1/m$.
		
\section{DHW formalism}\label{DHWformalism}
The research presented in this paper is based on the DHW formalism, a quantum kinetic method established in the phase space. In recent years, the DHW method has been widely applied in the study of electron-positron pair production \cite{DHW1,Blinne2016,DHW3,DHW4,DHW5,DHW6}. Its advantage lies in its ability to provide complete phase-pace information for particle production in a vacuum, and it is applicable under any given background field. A detailed derivation of the DHW formalism is provided in \cite{vortices1,Blinne2016}; therefore, we briefly introduce the ideas and key points of the DHW method below.
First, we write the Lagrangian in the QED as
\begin{equation}\label{1}
\begin{aligned}
L(\Psi ,\bar{\Psi },A)=\frac{1}{2}\left( \text{i}\bar{\Psi }{{\gamma }^{\mu }}{{D}_{\mu }}\Psi -\text{i}\bar{\Psi }D_{\mu }^{\dagger }{{\gamma }^{\mu }}\Psi  \right) -m\bar{\Psi }\Psi -\frac{1}{4}{{F}_{\mu \nu }}{{F}^{\mu \nu }},
\end{aligned}
\end{equation}
where ${{\mathcal{D}}_{\mu }}=\left( {{\partial }_{\mu }}+\text{i}e{{A}_{\mu }} \right)$ and $\mathcal{D}_{\mu }^{\dagger }=\left( \overleftarrow{{{\partial }_{\mu }}}-\text{i}e{{A}_{\mu }} \right)$ are the covariant derivatives with a vector potential ${{A}_{\mu }}$ that vanishes at asymptotic times, and ${{\gamma }^{\mu }}$ are the gamma matrices. Subsequently, using the Euler-Lagrange equation
 \begin{equation}\label{2}
		\begin{aligned}
{\partial _\mu }\left( {\frac{{\partial L}}{{\partial \left( {{\partial _\mu }{\rm{\bar \Psi }}} \right)}}} \right) - \frac{{\partial L}}{{\partial {\rm{\bar \Psi }}}} = 0,
\end{aligned}
	\end{equation}
we obtain the Dirac equation
\begin{equation}\label{3}
\begin{aligned}
\left( {{\rm{i}}{\gamma ^\mu }{\partial _\mu } - e{\gamma ^\mu }{A_\mu } - m} \right){\rm{\Psi }} = 0,~~~~~
{\rm{\bar \Psi }}\left( {{\rm{i}}{\overleftarrow{{{\partial }_{\mu }}} }{\gamma ^\mu } + e{\gamma ^\mu }{A_\mu } + m} \right) = 0.
\end{aligned}
\end{equation}
Starting from the field operator ${\rm{\Psi }}$, we constructed a gauge-covariant density operator in the Heisenberg picture:
\begin{equation}\label{4}
\begin{aligned}
{\hat {\cal C}_{\alpha \beta }}(r,s) = U(A,r,s)\left[ {{{{\rm{\bar \Psi }}}_\beta }(r - s/2),{{\rm{\Psi }}_\alpha }(r + s/2)} \right],
\end{aligned}
\end{equation}
where $r$ and $s$ are the center of mass and relative coordinates, respectively. To maintain the gauge invariance of the density operator, we introduce the Wilson line factor into the equation above:
\begin{equation}\label{5}
	\begin{aligned}
U(A,r,s) = \exp \left( {ies\int _{ - 1/2}^{1/2}{\rm{d}}\xi A(r + \xi s)} \right).
    \end{aligned}
\end{equation}
The Fourier transform of the density operator with respect to the relative coordinate $s$ yields the covariant Wigner operator
\begin{equation}\label{6}
	\begin{aligned}
{\hat {\cal W}_{\alpha \beta }}(r,p) = \frac{1}{2}\int {{\rm{d}}^4}s{{\rm{e}}^{{\rm{i}}ps}}{\hat {\cal C}_{\alpha \beta }}(r,s).
    \end{aligned}
\end{equation}
Taking the vacuum expectation value of the Wigner operator gives the Wigner function
\begin{equation}\label{7}
	\begin{aligned}
\mathbbm{W}(r, p)  =\langle\Phi|\hat{\mathcal{W}}(r, p) |\Phi\rangle.
    \end{aligned}
\end{equation}
Performing a spinor decomposition on the Wigner function yields 16 covariant Wigner coefficients.
\begin{equation}\label{8}
	\begin{aligned}
\mathbbm{W} = \frac{1}{4} \left( \mathbbm{1} \mathbbm{S} + \textrm{i} \gamma_5
		\mathbbm{P} + \gamma^{\mu} \mathbbm{V}_{\mu} + \gamma^{\mu} \gamma_5
		\mathbbm{A}_{\mu} + \sigma^{\mu \nu} \mathbbm{T}_{\mu \nu} \right).\
    \end{aligned}
\end{equation}
Here $\mathbb{S}$, $\mathbb{P}$, ${{\mathbb{V}}_{\mu }}$, ${{\mathbb{A}}_{\mu }}$ and ${{\mathbb{T}}_{\mu \nu }}$ are scalar, pseudoscalar, vector, axial-vector and tensor, respectively, where $\mathbb{S}$ is related to the mass density, $\mathbb{P}$ is related to the psuedoscalar condensate density, ${{\mathbb{V}}_{\mu }}$ is related to the net fermion current density, ${{\mathbb{A}}_{\mu }}$ is related to the polarization density, and ${{\mathbb{T}}_{\mu \nu }}$ is related to the electric dipole-moment density \cite{1991}. According to Refs. \cite{DHW1,1991,HebenstreitPHD}, the dynamical equation for the Wigner function is
\begin{equation}\label{9}
	\begin{aligned}
{{D}_{t}}\mathbb{W}=-\frac{1}{2}{{\mathbf{D}}_{\mathbf{x}}}\left[ {{\gamma }^{0}}{\bm{\gamma}} ,\mathbb{W} \right]+im\left[ {{\gamma }^{0}},\mathbb{W} \right]-i\mathbf{P}\left\{ {{\gamma }^{0}}{\bm{\gamma}} ,\mathbb{W} \right\}.
    \end{aligned}
\end{equation}
Here, ${D_t}$, ${{\bf{D}}_{\bf{x}}}$ and ${\bf{P}}$ denote pseudodifferential operators.
Because the electric field is spatially uniform, using the characteristic method\cite{Blinne2014} and represents momentum with canonical momentum ${\rm{\bf{p}  =  }}{\bf{q}} - e{\bf{A}}(t){\rm{\;}}$, which simplifies the 16 Wigner component partial differential equations into 10 ordinary differential equations. The nonvanishing Wigner coefficients are
\begin{equation}\label{10}
	\begin{aligned}
\mathbbm{w}=\left( \mathbbm{s},{{\mathbbm{v}}_{i}},{{\mathbbm{a}}_{i}},{{\mathbbm{t}}_{i}} \right),~~~~~{{\mathbbm{t}}_{i}}:={{\mathbbm{t}}_{0i}}-{{\mathbbm{t}}_{i0}.}
    \end{aligned}
\end{equation}
Because the equations for these $10$ Wigner coefficients are quite long, see  \cite{CKPHD,HebenstreitPHD} for the detailed form and derivation. The corresponding vacuum nonvanishing initial values are given by the following equation:
\begin{equation}\label{11}
	\begin{aligned}
{{\mathbbm{s}}_{vac}} = \frac{{ - 2m}}{{\sqrt {{{\bf{p}}^2} + {m^2}} }} , ~~~~~{{\mathbbm{v}}_{i,vac}}=\frac{-2{{p}_{i}}}{\sqrt{{{\mathbf{p}}^{2}}+{{m}^{2}}}}.
    \end{aligned}
\end{equation}
An important auxiliary quantity is introduced here: the phasespace energy density of the fermion field
\begin{equation}\label{12}
	\begin{aligned}
\varepsilon  = m\mathbbm{s} + {p_i}{{\mathbbm{v}}_{i}}.
    \end{aligned}
\end{equation}
Thus, the single-particle distribution function can be expressed as the difference between the energy density and vacuum energy density:
\begin{equation}\label{13}
	\begin{aligned}
f({\bf{q}},t) = \frac{1}{{2{\rm{\Omega }}({\bf{q}},t)}}\left( {\varepsilon  - {\varepsilon _{vac}}} \right),
    \end{aligned}
\end{equation}
where ${\rm{\Omega }}({\bf{q}},t) = \sqrt {{{\bf{p}}^2}(t) + {m^2}}  = \sqrt {{m^2} + {{[{\bf{q}} - e{\bf{A}}(t)]}^2}} $ denotes the total energy of particles. Expressing the above equation in terms of the projection of the Wigner function, and applying the Dirac equation, the following ordinary differential equation is obtained~\cite{Amat:2023vwv,Majczak:2024hmt,Aleksandrov:2024rsz,Amat:2024nvg}:
\begin{equation}\label{E19}
	\begin{array}{l}
	\displaystyle
	\dot{f}=\frac{e}{2{\rm{\Omega }}} \, \,  \mathbf{E}\cdot \mathbf{v},\\[2mm]
	\displaystyle
	\dot{\mathbf{v}}=\frac{2}{{\rm{\Omega }}^{3}}
	\left( (e\mathbf{E}\cdot \mathbf{p})\mathbf{p}-e{\rm{\Omega }}^{2}\mathbf{E}\right) (f-1)
	-\frac{(e\mathbf{E}\cdot \mathbf{v})\mathbf{p}}{{\rm{\Omega }}^{2}}
	-2\mathbf{p}\times \mathbbm{a} -2m \mathbbm{t},\\[2mm]
	\displaystyle
	\dot{\mathbbm{a}}=-2\mathbf{p}\times \mathbf{v},\\
	\displaystyle
	\dot{\mathbbm{t}}=\frac{2}{m}[m^{2}\mathbf{v}+(\mathbf{p}\cdot \mathbf{v})\mathbf{p}],
	\end{array}
\end{equation}
with the initial conditions $f({\bf{q}}, - \infty ) = 0,~{\bf{v}}({\bf{q}}, - \infty ) = \mathbbm{a}({\bf{q}}, - \infty ) = \mathbbm{t}({\bf{q}}, - \infty ) = 0 $. where the dots represent time derivatives and ${\bf{E}}$ is the electric field of Eq. (\ref{fieldmode3}), and $e$ is the charge of particle, ${\bf{A}}(t)$ is the vector potential of the external field. Finally, the number density of the resulting electron--positron pairs can be obtained by integrating the distribution function $f({\bf{q}},t)$ over the full momentum at $t \to  + \infty $.
\begin{equation}\label{14}
	\begin{aligned}
n=\underset{t\to +\infty }{\mathop{\lim }}\,\int \frac{{{d}^{3}}q}{{{(2\pi )}^{3}}}f(\mathbf{q},t).
    \end{aligned}
\end{equation}

In addition, the above distribution function does not distinguish the spin states of the particles. When we take the spin effect into account, we can extract information regarding the spin and chiral signatures of the resulting electron-positron pairs from the Wigner function, and the spin dependent single-particle momentum distribution function should change to \cite{Blinne2016,hulinaspin}
\begin{equation}\label{15}
	\begin{aligned}
{f_s} = \frac{1}{2}\left( {f + s\delta {f_{sc}}} \right),
    \end{aligned}
\end{equation}
where $s =  \pm 1$, $\delta {{f}_{sc}}=\frac{{{q}_{z}}}{{{\epsilon }_{\bot }}}\delta {{f}_{\text{c}}}+\frac{m}{{{\epsilon }_{\bot }}}\delta {{f}_{{{\mu }_{\text{z}}}}}$, the chiral asymmetry $\delta {f_{\rm{c}}}$ can be defined as $\delta {{f}_{\text{c}}}=\frac{1}{2\text{ }\!\!\Omega\!\!\text{ }}\mathbf{p}\cdot \mathbbm{a}$, and the magnetic moment asymmetry $\delta {{f}_{{{\mu }_{z}}}}=\frac{1}{2\text{ }\!\!\Omega\!\!\text{ }}\left( m{{a}_{z}}+{{(\mathbf{p}\times \mathbbm{t})}_{z}} \right)$, ${{\epsilon }_{\bot }}=\sqrt{{{m}^{2}}+q_{z}^{2}}$. The corresponding number density can also be modified as
\begin{equation}\label{16}
	\begin{aligned}
{{n}_{s}}=\underset{t\to +\infty }{\mathop{\lim }}\,\int \frac{{{d}^{3}}q}{{{(2\pi )}^{3}}}{{f}_{s}}(\mathbf{q},t).
    \end{aligned}
\end{equation}

\section{Numerical results}\label{results}
\subsection{Momentum spectrum without spin effect}
\begin{figure*}[ht!]
\centering\includegraphics[width=1\linewidth]{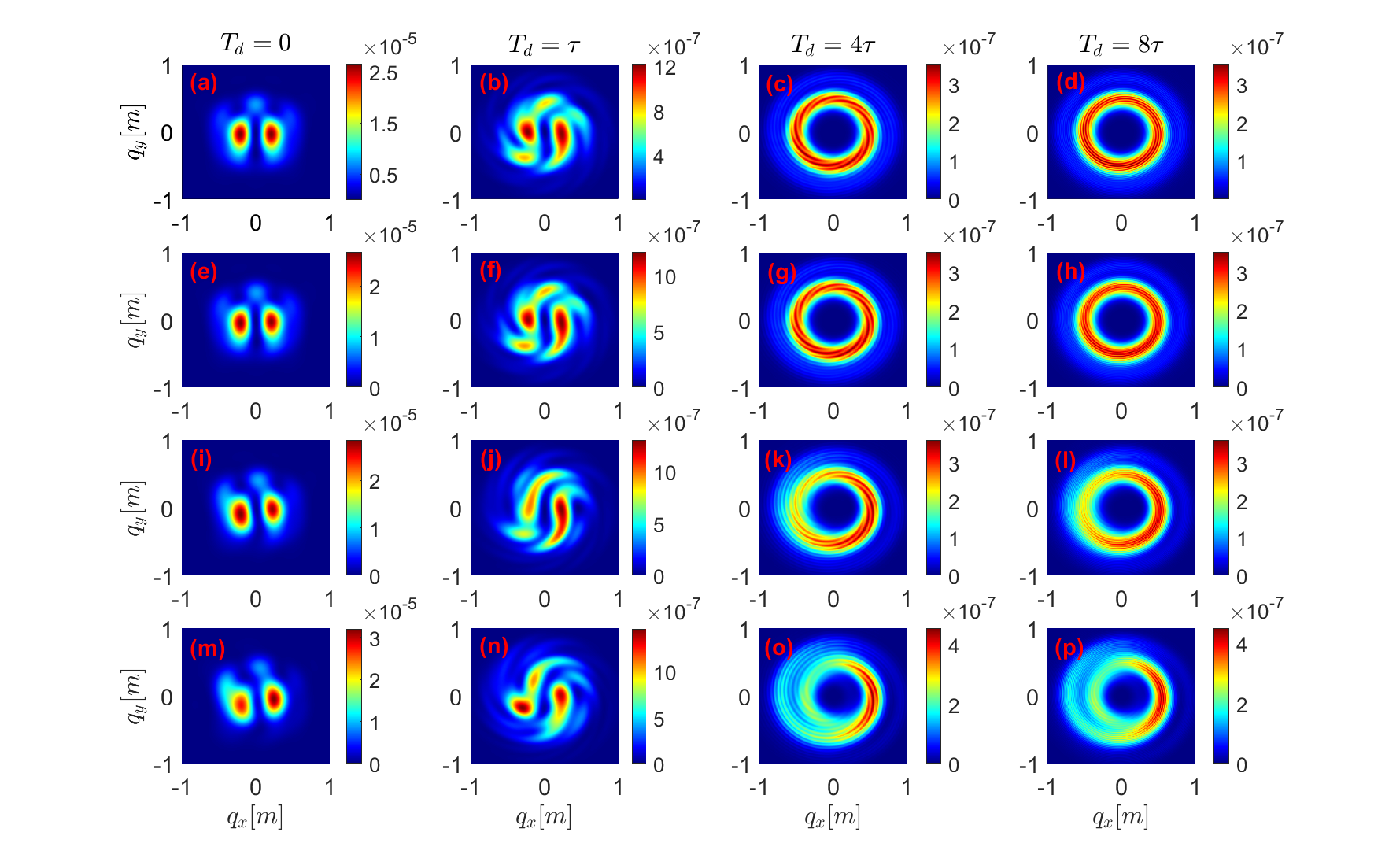}
\caption{The momentum spectrum of particles in the $q_x-q_y$ plane (where $q_z=0$) for a symmetric electric field ($k=1$) with different time delay and frequency chirp. From left to right in each column, the time delay parameters are ${T_d} = 0$, ${T_d} =~\tau$, ${T_d} = 4\tau$, and ${T_d} = 8\tau$, respectively; from top to bottom in each row, the frequency chirp parameters are $b = 0$, $b=0.01\omega /\tau$, $b=0.05\omega /\tau$ and $b=0.1\omega /\tau$, respectively. The other field parameters are given in introduction text for Eq. (\ref{fieldmode3}).}
\label{fig1}
\end{figure*}
\begin{figure*}[ht!]
\centering\includegraphics[width=1\linewidth]{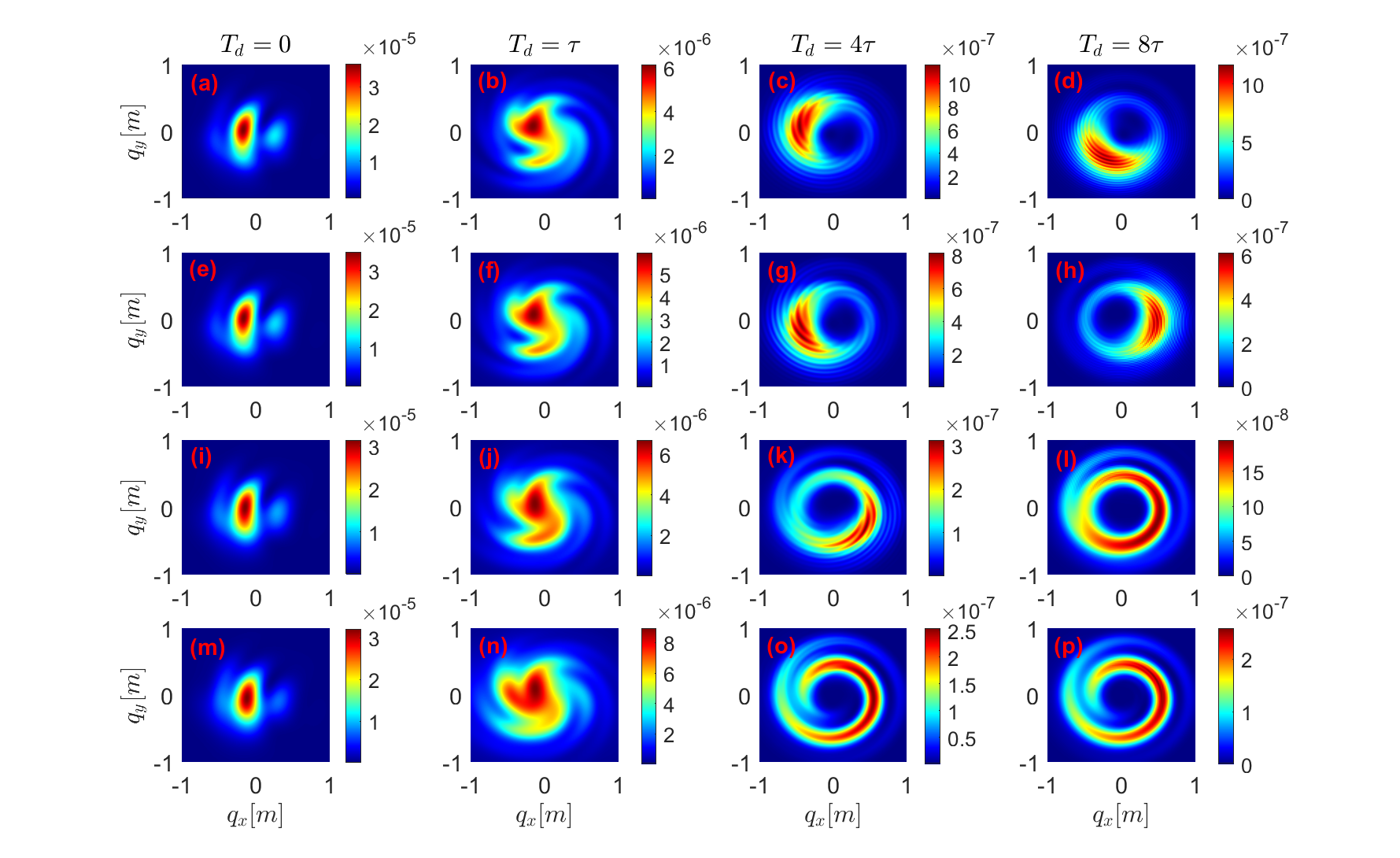}
\caption{The momentum spectrum of particles in the $q_x-q_y$ plane (where $q_z=0$) for a shortened asymmetric electric field ($k=0.3$) with different time delay and frequency chirp. From left to right in each column, the time delay parameters are ${T_d} = 0$, ${T_d} = \tau$, ${T_d} = 4\tau$ and ${T_d} = 8\tau$, respectively; from top to bottom in each row, the frequency chirp parameters are $b = 0$, $b=0.01\omega /\tau$, $b=0.05\omega /\tau$ and $b=0.1\omega /\tau$, respectively. The other field parameters are given in introduction text for Eq. (\ref{fieldmode3}).}
\label{fig2}
\end{figure*}
\begin{figure*}[ht!]
\centering\includegraphics[width=1\linewidth]{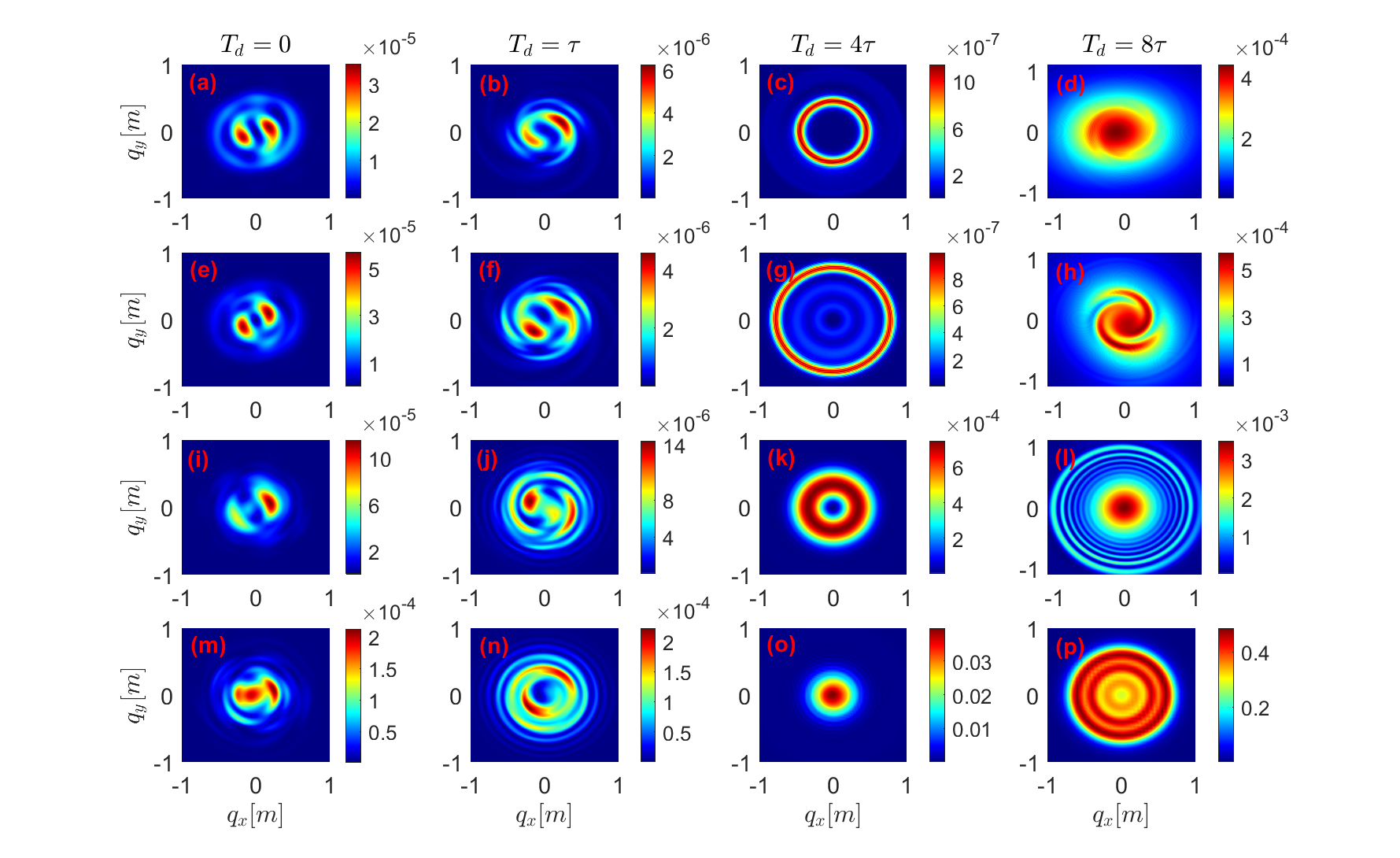}
\caption{The momentum spectrum of particles in the $q_x-q_y$ plane (where $q_z=0$) for an extended asymmetric electric field ($k=3$) with different time delay and frequency chirp. From left to right in each column, the time delay parameters are ${T_d} = 0$, ${T_d} = \tau$, ${T_d} = 4\tau$, and ${T_d} = 8\tau$, respectively; from top to bottom in each row, the frequency chirp parameters are $b = 0$, $b=0.01\omega /\tau$, $b=0.05\omega /\tau$ and $b=0.1\omega /\tau$, respectively. The other field parameters are given in introduction text for Eq. (\ref{fieldmode3}).}
\label{fig3}
\end{figure*}
In the previous studies \cite{vortices2,vortices3,vortices4,hulinaspin}, the effect of the time delay on the momentum spectrum of particles generated in two counter-rotating electric fields was discussed. Momentum spirals are sensitive to the time delay within the fields. An increase in the delay time causes the spiral arms to become thinner and longer, and the number of spiral arms increases significantly. A small time delay can lead to a significant increase in the particle number density by a factor of approximately five. With the introduction of the frequency chirp and field asymmetry, do the above conclusions remain consistent? Therefore, based on the results presented above, in this work we will focus on a combined effect of time delay, field asymmetry and frequency chirp on the momentum spectrum, as shown in Figs. \ref{fig1}, \ref{fig2} and \ref{fig3}. Note that, in order to prove the validity of our field configuration, we considered the simple case of time-delayed symmetric field without chirp, which is considered previous researches mentioned above, and the results are compared with combined effects in our case.

For a symmetric electric field ($k=1$) without chirp and time delay ($b = 0$ and ${T_d}=0$),the momentum spectrum in the ($q_x-q_y$) plane (polarization plane) for $q_z=0$ is shown in Fig. \ref{fig1} (a). The momentum spectrum peak is divided into two parts located in the positive and negative ${q_x}$ regions and is symmetric about the ${q_x}$ direction. This is mainly because without a time delay, the vector potential of the electric field ${A_x}(t)$ is an odd function of $t$ \cite{vortices4}. As the frequency chirp increased, the particle momentum spectrum changed significantly, with a broken symmetry of the distribution. This may be related to post dynamics of created pairs on the asymmetric electric field. The broken symmetry at the time of the chirped field causes different particle dynamics after creation. This shows that the peak of the momentum spectrum shifts towards the positive ${q_x}$ region with an increase in the chirp parameter $b$, as shown in Figs. \ref{fig1} (e) (i) and (m). For the unchirped case ($b=0$), as the time delay $T_d$ increased, momentum spectrum spirals appeared as shown in Figs. \ref{fig1} (a)-(d), and the particle number density decreased by approximately two orders of magnitude. Notably, in the case of $T_d=4\tau$ and $b=0.01\omega/\tau$ the vortex structure of the momentum spectra is more obvious, but the symmetry of the spectra is violated with an increase in $b$ for a fixed $T_d$. Moreover, the interference effects become visible with increasing $T_d$ and $b$ for a symmetric electric field $k=0$. The interference pattern can be interpreted as the result of interference between two temporally separated pair creation events, coupled with the complex oscillation induced by a large chirp.  Electrons created at different times exhibit a non-vanishing relative phase, leading to interference during the pair creation process.  This relative phase is momentum-dependent, with momentum being influenced by the electric field.  Consequently, this results in the interference transitioning between constructive and destructive states, thereby causing characteristic variations in the momentum spectra that are notably relying on form of the applied electric field.

For shorter asymmetric electric field with $k=0.3$, the temporal asymmetry of the electric field results in an increased asymmetry in the momentum spectrum distribution, as shown in Fig. \ref{fig2}. The results for the particle number density in the momentum spectrum are consistent with those of the symmetric field case, and for large time delays, the particle number density also decreases. The asymmetric electric field caused a shift in the momentum peak, and the time delay between the two fields led to a vortex structure in the momentum spectrum. Modulation of the electric field by frequency chirp changes the frequency distribution of the field, causing the peak of the momentum spectrum of the generated particles to change accordingly. Under the combined influence of these factors, the peak of the particle momentum spectrum demonstrates a rotational trend as the frequency chirp increases. However, this can also be observed in Fig. \ref{fig2} that under larger chirp parameter $b$, as the time delay increases, there is only one long and thick spiral arm in the momentum spectrum. Compared to the symmetric electric field case (see Fig. \ref{fig1}), the interference fringes vanish with the increases of the $k$, and the distinction between the spiral arms becomes more pronounced. The results indicate that in the case of an asymmetric electric field with a larger time delay and chirp parameter, the momentum vortex may be enhanced.

For extended asymmetric electric field ($k = 3$), there was a significant difference in the change of the particle number density under smaller- and larger-frequency chirp parameters. When the frequency chirp is small and there is no time delay, the particle number density increases slightly compared to the symmetric ($k=1$) and shorter asymmetric ($k=0.3$) electric fields; when the frequency chirp is large and there is no time delay, the particle number density significantly enhances, as shown in Fig. \ref{fig3} (m). This phenomenon can be attributed to the extended field length encompassing a greater number of photons. A larger chirp parameter corresponds to a higher effective frequency, leading to a more intricate electric field structure that facilitates enhanced particle production. When we increased the time delay, the trend of the momentum spectrum distribution was completely different compared to the symmetric and shorter asymmetric field cases. For smaller frequency chirp parameter ($b=0.01\omega/\tau$), as the time delay increases, the momentum spectrum gradually exhibits a vortex shape (Fig. \ref{fig3} (f)), and then transitions into a ring structure, as seen in Fig. \ref{fig3} (g), with the number density gradually decreasing. When the time delay increases to $T_d=8\tau$, the peak of the momentum spectrum shifts to the central area, displaying two spiral arms, and the number density increases significantly, as shown in Fig. \ref{fig3} (h). For larger chirp parameter ($b=0.1\omega/\tau$), when $T_d=\tau$ an obvious vortex structure is observed in the central region surrounded by low-density rings, as shown in Fig. \ref{fig3} (n). As the time-delay increases to $T_d=4\tau$ the momentum spectrum shrinks to the center, and the number density is enhanced by two orders of magnitude, as shown in Fig. \ref{fig3} (o). Finally, for $T_d=8\tau$, the number density was continually enhanced and the momentum distribution exhibited a multiphoton ring structure, as shown in Fig. \ref{fig3} (p). It is obvious that in the case of $k=3,~T_d=8\tau,~b=0.1\omega/\tau$ the number density increases by six orders of magnitude compared to the symmetric electric field case ($k=1,~T_d=8\tau,~b=0.1\omega/\tau$), as shown in Fig. \ref{fig1} (p). This suggests that an extended asymmetric rotating electric field with a large-frequency chirp results in a larger field extent and more rapid oscillation. Thus, a time delay may enhance certain specific multiphoton processes. The above results indicate that in the case of extended electric fields, not only the large-frequency chirp but also the large time delays can enhance the number density of particles.

\subsection{Spin dependent momentum spectrum}
\begin{figure*}[ht!]
\centering\includegraphics[width=1\linewidth]{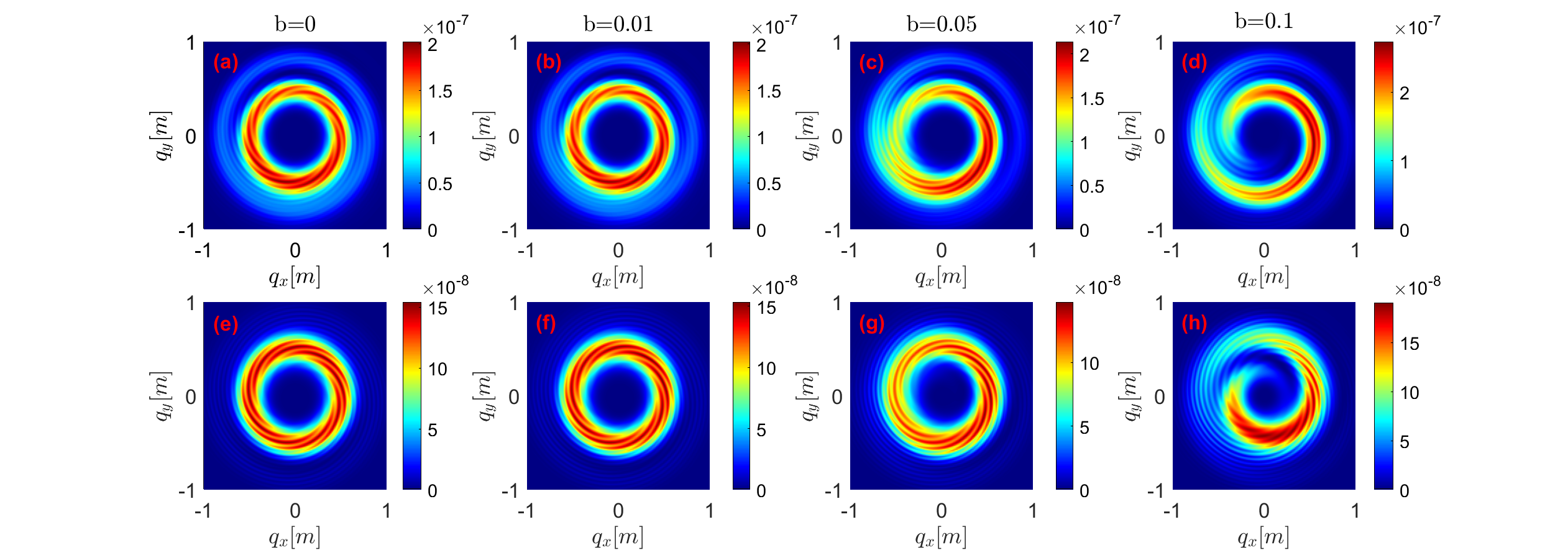}
\caption{The momentum spectrum of particles in the $q_x-q_y$ plane (where $q_z=0$) for a symmetric electric field ($k=1$) with frequency chirp. From top to bottom, each row represents the momentum spectrum for ${f_ + }\left( {{q_x},{q_y}} \right)$ and ${f_ - }\left( {{q_x},{q_y}} \right)$, respectively. From left to right, the spectra correspond to the cases where $b = 0$, $b=0.01\omega /\tau$, $b=0.05\omega /\tau$ and $b=0.1\omega /\tau$, respectively. The time delay between the two fields is fixed to ${T_d} = 4\tau $. The other field parameters are given in introduction text for Eq. (\ref{fieldmode3}).}
\label{fig4}
\end{figure*}	
\begin{figure*}[ht!]
\centering\includegraphics[width=1\linewidth]{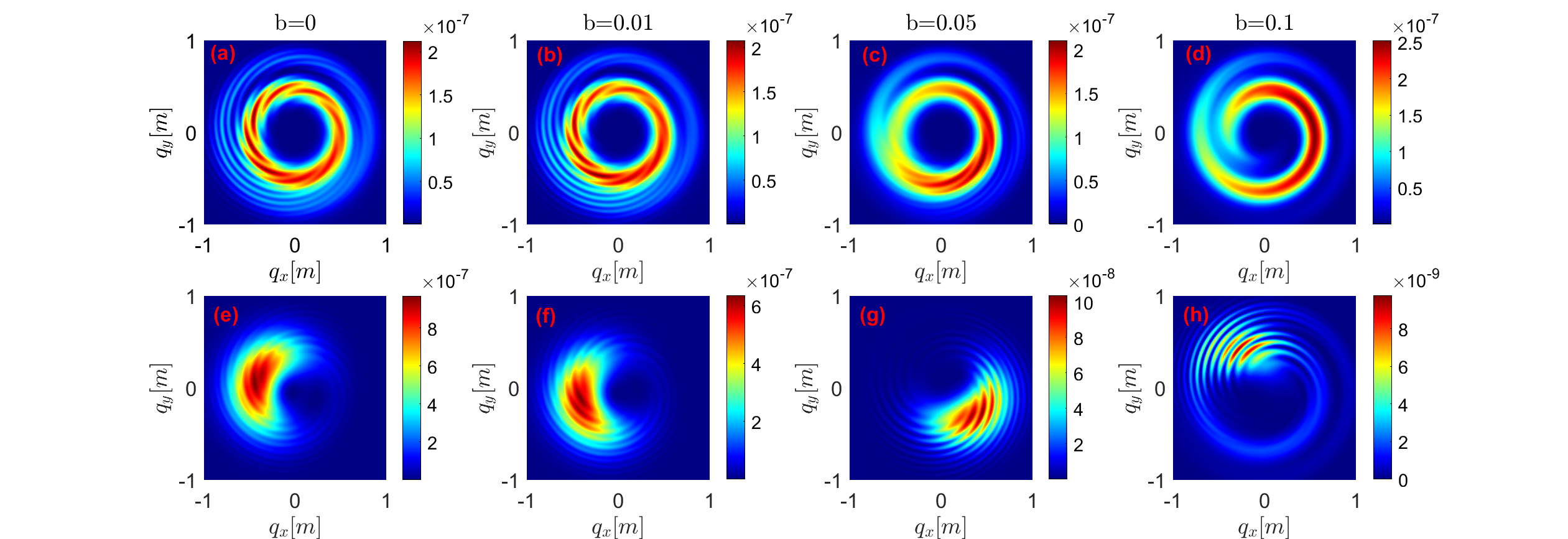}
\caption{The momentum spectrum of particles in the $q_x-q_y$ plane (where $q_z=0$) for an asymmetric electric field ($k=0.3$) with frequency chirp. From top to bottom, each row represents the momentum spectrum for ${f_ + }\left( {{q_x},{q_y}} \right)$ and ${f_ - }\left( {{q_x},{q_y}} \right)$, respectively. From left to right, the spectra correspond to the cases where $b = 0$, $b=0.01\omega /\tau$, $b=0.05\omega /\tau$ and $b=0.1\omega /\tau$, respectively. The time delay between the two fields is fixed to ${T_d} = 4\tau $. The other field parameters are given in introduction text for Eq. (\ref{fieldmode3}).}
\label{fig5}
\end{figure*}
\begin{figure*}[ht!]\suppressfloats
\centering\includegraphics[width=1\linewidth]{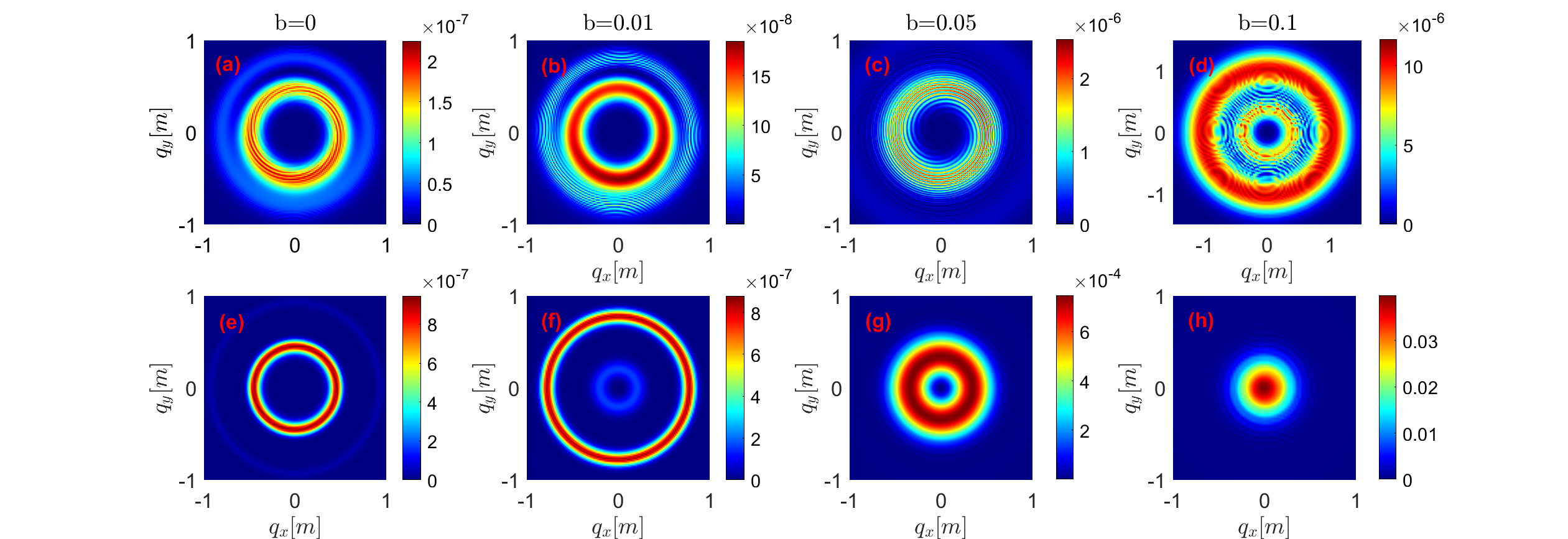}
\caption{The momentum spectrum of particles in the $q_x-q_y$ plane (where $q_z=0$) for an asymmetric electric field ($k=3$) with frequency chirp. From top to bottom, each row represents the momentum spectrum for ${f_ + }\left( {{q_x},{q_y}} \right)$ and ${f_ - }\left( {{q_x},{q_y}} \right)$, respectively. From left to right, the spectra correspond to the cases where $b = 0$, $b=0.01\omega /\tau$, $b=0.05\omega /\tau$ and $b=0.1\omega /\tau$, respectively. The time delay between the two fields is fixed to ${T_d} = 4\tau $. The other field parameters are given in introduction text for Eq. (\ref{fieldmode3}).}
\label{fig6}
\end{figure*}
Particle-pair production is a complex process, and the number density and momentum spectra of the created pairs are highly sensitive to changes of the background fields.
When pair creation in a standing wave field is studied, neglecting the magnetic field component and the spatial variation of the electric field is a commonly applied approximation.  The pair production process is modified if the direction of the electric field changes over time even in pure electric fields. In spatially homogeneous electric fields, the time dependence results in enhancement of pair creation and quantum interference, which distorts momentum spectra.
In the previous section, the momentum spectra of the pairs in different field configurations under the standing wave approximation are studied in detail.
Generally, the pair production process is spin dependent and both the spin-up and spin-down states contribute to the momentum spectrum and sensitive to field parameters \cite{hulinaspin}. Moreover, for a purely time-varying electric field with spatial homogeneity and a vanishing magnetic field, the vacuum pair production process is also spin-dependent because of the Lorentz boost in the electron's rest frame and the spin-orbit coupling inherent in the Dirac equation \cite{spin8}. For a rotating electric field, the spin state in the momentum spectrum of the created pairs depends on the ranges of electric field lengths and frequencies. In this regard, it is interesting to discuss the particles spin dependent momentum spectrum as a function of field asymmetry parameter $k$, time delay $T_d$ and chirp parameter $b$.	

Figures \ref{fig4}-\ref{fig6} show the spin-dependent momentum spectra of the particles produced in the two-color counter-rotating frequency-chirped asymmetric electric fields. For the spin-up and spin-down distribution functions ${f_ \pm }\left( {{q_x},{q_y}} \right)$, the momentum distributions exhibited significant differences for different field parameters.
The spin-considered momentum spectrum of the particles in a symmetric ($k=1$) field with various frequency chirps for a fixed delay time $T_d=4\tau$ is shown in Fig. \ref{fig4}. The difference between spin-up ${f_ + }\left( {{q_x},{q_y}} \right)$ and spin-down ${f_ - }\left( {{q_x},{q_y}} \right)$ distribution is small, and the ${f_ + }\left( {{q_x},{q_y}} \right)$ having a slightly higher peak density as shown in Figs. \ref{fig4} (a) and (e). As the chirp parameter $b$ increases, both the spin-up and spin-down pair densities increase slightly, which is consistent with the discussion in the previous section.
The larger chirp gives larger effective frequency, then the total number density increases with the chirp parameter $b$ (see Fig. \ref{fig1}) in symmetric electric field. For the distribution pattern of the momentum spectrum, because the time variation of the frequency is asymmetric within the symmetric field envelope, this results different particle dynamics after creation. Then the symmetry of the momentum spectrum is broken with an increasing chirp parameter $b$, and the spiral arms branch off more clearly in the spin-down spectrum for a larger $b$. However, the momentum distributions of the spin-up particles gradually transform from being composed of multiple thin, elongated spiral arms to a single thicker spiral arm as the chirp parameter increases.

Subsequently, we examined the effects of the field asymmetry with frequency chirp on the spin-dependent momentum spectrum. For a small degree of field asymmetry parameter ($k=0.3$), the difference between the distributions of ${f_ + }\left( {{q_x},{q_y}} \right)$ and ${f_ - }\left( {{q_x},{q_y}} \right)$ is pronounced, as shown in Fig. \ref{fig5}. In the absence of chirp ($b=0$) and with small chirp ($b=0.01\omega/\tau$), the number density of the spin-down momentum spectrum ${f_ - }\left( {{q_x},{q_y}} \right)$ is three to four times higher than that for spin-up momentum spectrum ${f_ + }\left( {{q_x},{q_y}} \right)$, indicating that particle production is dominated by spin-down particles in this case, as shown in Figs. \ref{fig5} (a), (b), (e), and (f).
Alternatively, the spin dependence can be understood based on symmetry considerations. In the case of asymmetric electric fields, the direction of rotation reverses upon reflection about the z-axis, which implies that the probability of pair creation may exhibit spin dependence.
For the unchirped or very small chirped case, for example, electrons are created preferably with spin down, which propagate only along the negative $z$ direction, and with spin down, which exhibits a complete vortex patten.
However, with increases of the frequency chirp to $b=0.05\omega/\tau$, the peak of the momentum spectrum for spin-down distribution ${f_-}\left( {{q_x},{q_y}} \right)$ rapidly decreases and starts to be lower than that spin-up case ${f_ + }\left( {{q_x},{q_y}} \right)$, and it significantly decreases with further increase with chirp parameter $b$. When the chirp parameter reached $b=0.1\omega/\tau$, the number density decreased by one order of magnitude, and particle production was then dominated by spin-up particles, as shown in Figs. \ref{fig5} (d) and (h). Obviously, under the combined effect of the field asymmetry and frequency chirp, for the
asymmetric electric field case with $k=0.3$, the spin asymmetry degree reversed as the chirp parameter increased. This will be discussed in detail below.

For an extended asymmetric field shape with a larger degree of asymmetry,  $k=3$, Fig. \ref{fig6} presents the distribution of the  spin-up ${f_ + }\left( {{q_x},{q_y}} \right)$ and spin-down ${f_ - }\left( {{q_x},{q_y}} \right)$ momentum spectrum. Compared to the results shown in Figs. \ref{fig4} and \ref{fig5}, we find that regardless of the presence of frequency chirp and the magnitude of the chirp parameter, ${f_ - }\left( {{q_x},{q_y}} \right)$ always dominates. The number density of the particles in ${f_ + }\left( {{q_x},{q_y}} \right)$ increases with an increase in the frequency chirp, but the rate of increase is much slower than that in the spin-down distribution ${f_ - }\left( {{q_x},{q_y}} \right)$. For unchirped case, the spin-up particle density in ${f_ + }\left( {{q_x},{q_y}} \right)$ is four times lower than that in the spin-down distribution ${f_ - }\left( {{q_x},{q_y}} \right)$ as shown in Figs. \ref{fig6} (a) and (e), and it differs by three orders of magnitude from ${f_ - }\left( {{q_x},{q_y}} \right)$ at larger chirp values $b=0.1\omega/\tau$ as shown in Figs. \ref{fig6} (d) and (h). It was also found that multi spiral arms are present and obvious interference fringes are observed in the spin-up momentum spectrum ${f_ + }\left( {{q_x},{q_y}} \right)$ whereas the distribution changes from multi-photon ring to a spot-like distribution in the spin-down spenctrum ${f_-}\left( {{q_x},{q_y}} \right)$, as shown in Figs. \ref{fig6} (e)-(h).
\begin{figure}[ht!]
\centering\includegraphics[width=1.1\linewidth]{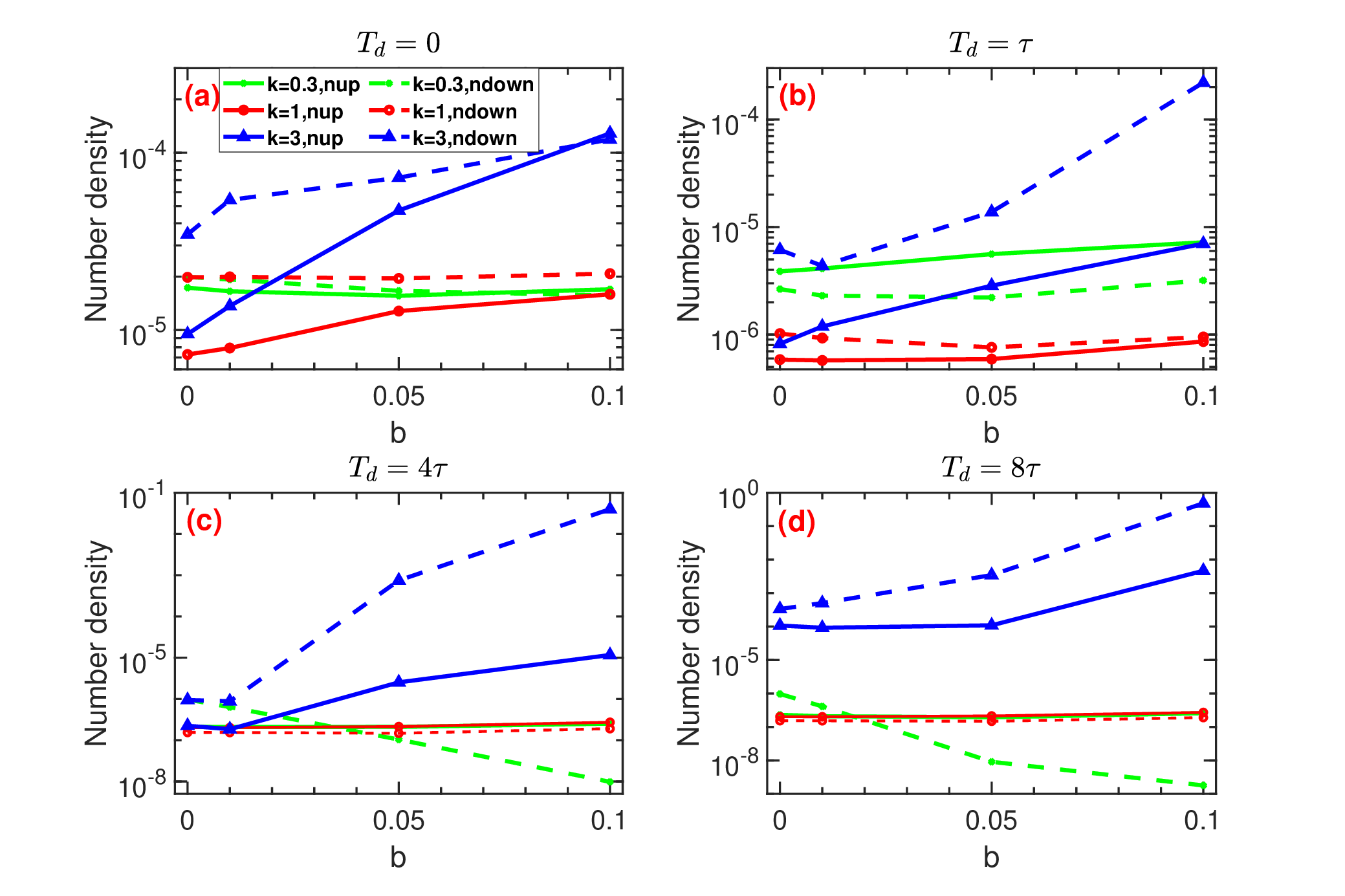}
\caption{The evolution of the spin-up (${n_+}$) and spin-down (${n_ - }$) particle densities with the chirp parameter $b$ for different time delay $T_d$; (a) for $T_d=0$, (b) for $T_d=\tau$, (c) for $T_d=4\tau$ and (d) for $T_d=8\tau$, respectively. The red curves for symmetric electric field ($k=1$), and the green and blue curves for asymmetric electric fields with  $k=0.3$ and $k=3$, respectively. The solid and dashed lines are represent to the spin-up (${n_ + }$) and spin-down (${n_ - }$) particle density. The other field parameters are
given in introduction text for Eq. (\ref{fieldmode3}).}
\label{fig7}
\end{figure}

In the combined field with different time delays, the influence of field asymmetry and frequency chirp on the particle number density is shown in Fig. \ref{fig7}. To clearly demonstrate the spin effect in pair production, we define the spin asymmetry degree on the number density of created particles as \cite{hulinaspin}:
 \begin{equation}
\kappa_n={\kappa_n^+}-{\kappa_n^-},
\end{equation}
where ${\kappa_n^+}=n_{+} /\left(n_{+}+n_{-}\right)$, $\kappa_n^-=n_{-} /\left(n_{+}+n_{-}\right)$ , and $ n_{+} $, $ n_{-} $ represent the number densities of spin-up and spin-down particles, respectively. Note that if ${\kappa_n^-} > {\kappa_n^+}$, then the spin asymmetry degree $\kappa_n$ is negative, which indicates that spin-down particles dominate in the pair production, and vice versa.

For symmetric electric field ($k=1$), for the cases ${T_d} = 0$ and ${T_d} = \tau $, the density of the spin-down particles ${n_ - }$ is always slightly greater than that of the spin-up particles ${n_ + }$ with increasing $b$ as shown in Figs. \ref{fig7} (a) and (b). However, as the time delay parameter increases to ${T_d} = 4\tau$ and $T_d = 8\tau$, the spin-up particle number ${n_ + }$ becomes slightly greater than ${n_ - }$, and ${n_ + }$ always dominates the pair production process, as shown in Figs. \ref{fig7} (c) and (d). For the case of vanishing chirp ($ b = 0 $), the spin asymmetry degree $ \kappa_n $ varies from $-46.31\%$ for $T_d = 0 $ to $ 13.61\%$ for $T_d = 8\tau $ as the time delay increases. Similarly, for other chirp parameters, the spin asymmetry degree $ \kappa_n $ also evolves from negative to positive values with increasing time delay. This indicates that the spin asymmetry degree also changes when the time delay increases. This result is consistent with the findings in \cite{hulinaspin}. For the shortened asymmetric electric field with $ k = 0.3 $, a reversal of the spin asymmetry degree can also be observed under varying time delays, wherein the spin asymmetry degree transitions from positive to negative or vice versa. For the complete superposition case (${T_d} = 0$), the reversal of the spin asymmetry degree occurs with a relatively large $b$, as shown in Fig. \ref{fig7} (a). For the time-delay parameter ${T_d} = \tau $, ${n_ + }$ is greater than ${n_ - }$ as shown in Fig. \ref{fig7} (b). However, for larger time delay parameters $ T_d = 4\tau $ and $ T_d = 8\tau $, when the chirp parameter $ b $ is relatively small (e.g., $ b =0.01$, $ T_d = 8\tau $), $ n_- $ exceeds $ n_+ $, resulting in a negative spin asymmetry degree ($ \kappa_n = -32.52\% $). As $ b $ increases, particularly for larger chirp parameters (e.g., $ b = 0.1\omega /\tau $, $ T_d = 8\tau $), $ n_+ $ surpasses $ n_- $, thereby reversing the spin asymmetry degree to a positive value ($ \kappa_n = 98.62\% $), as shown in Fig. \ref{fig7} (c) and (d). For extended asymmetric electric field $k=3$, both ${n_ +}$ and ${n_ - }$ increase with the chirp parameter $b$ and the time delay $T_d$. It is observed that the reversal of the spin asymmetry degree occurs exclusively when the time delay parameter $ T_d = 0 $, provided that the chirp parameter $ b $ is sufficiently large. Conversely, as the time delay increases, $ n_- $ consistently exceeds $ n_+ $ and remains dominant throughout. When both the time delay ($ T_d = 4\tau $) and the chirp parameter ($ b = 0.1\omega /\tau $) are set to relatively high values, the spin asymmetry degree $ \kappa_n $ reaches a value of $-99.94\%$.

Based on the above results, it was found that not only time delay but also frequency chirp can lead to a reversal of the spin asymmetry degree in the pair production process. Moreover, the outcomes exhibit variations depending on the characteristics of different asymmetric electric fields. This provides an alternative way to control the momentum spectrum and spin asymmetry degree of created particles by tuning the field asymmetry degree and the frequency chirp.
		
\section{Discussion and Conclusion}\label{conclusion}
In this study, we used the DHW formalism to investigate the effects of asymmetric electric field shapes, frequency chirps, and time delays on the momentum spectra of electron--positron pairs produced in a two-color counter-rotating electric fields. By analyzing the momentum spectra of the produced particles, we obtained the following results: the momentum spectrum of the produced pairs is highly sensitive to changes in the field asymmetry, frequency chirp, and time delay. The field asymmetry and frequency chirp lead to an asymmetric momentum distribution and changes in the particle number density. For the extended asymmetric electric fields with a large-frequency chirp parameter, the particle number density increased by more than four orders of magnitude. For symmetric and asymmetric fields, the effects of frequency chirp and time delay on the momentum spectrum of the produced particles were also different. For symmetric and shortened asymmetric electric fields, increasing the time delay to ${T_d} = 8\tau$ the particle number density is reduced by two orders of magnitude. For extended asymmetric fields, a large time delay (${T_d} = 8\tau$) increases the particle number density, and if the frequency chirp parameter is also large ($b=0.1\omega /\tau$), the particle number density increases by more than three orders of magnitude compared to the case without time delay; compared to the symmetric electric field without time delay and frequency chirp, the particle number density increases by six orders of magnitude.

A remarkable result of this paper is the observation that, by precisely adjusting the shape, frequency chirp, and time delay of the two rotating electric fields, spin-polarized pairs can be created. We studied the spin-dependent momentum spectra of ${f_ + }\left( {{q_x},{q_y}} \right)$ and ${f_ - }\left( {{q_x},{q_y}} \right)$, and found that both variations in time delay and changes in frequency chirp can lead to a reversal of the spin asymmetry degree. The outcomes differ depending on the shapes of the electric field. For symmetric electric field with no time delay (${T_d} = 0$) and a small time delay (${T_d} = \tau$), ${n_ - }$ dominated, whereas for a larger time delay (${T_d} = 4\tau, 8\tau$), ${n_ + }$ dominated. For shortened asymmetric electric fields ($k=0.3$), variations in both time delay and frequency chirp can induce a reversal of the spin asymmetry degree. However, for extended asymmetric electric fields ($k = 3$), reversal of the spin asymmetry degree occurs only under complete superposition of the two fields ($T_d = 0$) with a frequency chirp parameter ranging from $b = 0.05\omega/\tau$ to $b = 0.1\omega/\tau$. The spin asymmetry degree reaches 98.62\% with a large time delay ($T_d = 8\tau$) and large frequency chirp ($b = 0.1\omega/\tau$) for shortened asymmetric electric fields, while it reaches -99.94\% for extended asymmetric electric fields with $T_d = 4\tau$ and $b = 0.1\omega/\tau$. Although we have considered a slightly more complex case of the multi-parameter combination in this paper, the results cover the single-parameter cases and help to deepen the understanding of the parameter relation characteristics of the spin asymmetry effect in pair production.

It should be noted that in the absence of an external magnetic field, the Schwinger effect still exhibits spin dependence.
This spin dependence originates from the spin-orbit coupling term $\mathbf{s\cdot(p\times E)}$ in the Dirac equation, which is an inherent characteristic of relativistic effects.
Intuitively, the electrons generated from the vacuum move at a velocity $\mathbf{v}$ under the action of the electric
field $\mathbf{E}$.
In their rest frame, the electrons will feel an effective magnetic field $\mathbf{B_{eff}\sim v\times E}$.
Therefore, due to the Zeeman coupling $\mathbf{s\cdot B_{eff}}$, the single-particle energy will have spin dependence.
For the rotating electric field in our case, the direction of the electric field $\mathbf{E}$ changes with time, the electron's velocity $\mathbf{v}$ is no longer parallel to $\mathbf{E}$, causing the spin to align along the $\mathbf{v\times E}$ direction. This process can alter the momentum spectrum,
making the particle's momentum $\mathbf{p}$ and spin $\mathbf{s}$ direction tend to be parallel. This can be attributed to the relativistic effect, originating from the spinor conservation in the relativistic limit.

Therefore, it would be interesting to study the spin effects through time-dependent rotating fields and standing wave approximations without magnetic field. Our findings suggest a novel viewpoint: spin, to further study the vacuum pair production, and propose novel spin-dependent observables, which are not only interesting to high-energy particle physics, but also to condensed matter and material physics.

\begin{acknowledgments}
We are grateful to the anonymous referee for critical reading and helpful suggestions to improve the manuscript. This work was supported by the National Natural Science Foundation of China (NSFC) under Grant No. 12265024 and the Special Training Program of the Science and Technology Department of Xinjiang China (Grant No. 2024D03007).
O. Amat acknowledges support from the National Natural Science Foundation of China (NSFC) under Grant No.12447179.
\end{acknowledgments}


\begin{thebibliography}{apsrev4-1}

\bibitem{Sauter:1931zz}
F.~Sauter,
Uber das Verhalten eines Elektrons im homogenen elektrischen Feld nach der relativistischen Theorie Diracs,
\href{https://link.springer.com/article/10.1007/BF01339461}{Z. Phys. \textbf{69}, 742 (1931).}
	
\bibitem{Schwinger:1951nm}
J.~S.~Schwinger,
For gauge invariance and vacuum polarization,
\href{https://journals.aps.org/pr/abstract/10.1103/PhysRev.82.664}{Phys. Rev. \textbf{82} and 664 (1951).}


\bibitem{Strickland:1985gxr}
D.~Strickland and G.~Mourou,
Compression of amplified chirped optical pulses,
\href{https://www.sciencedirect.com/science/article/abs/pii/0030401885901518?via3Dihub}{Opt. Commun. \textbf{55}, 447-449 (1985).}


\bibitem{eli-beams}
\href{http://www.eli-beams.eu}{http://www.eli-beams.eu.}

\bibitem{xcels}
\href{http://xcels.iapras.ru}{http://xcels.iapras.ru.}

\bibitem{Keldysh:1965ojf}
L.~V.~Keldysh,
Ionization in the field of strong Electromagnetic waves,
\href{https://inspirehep.net/files/6697e05d52e411291acc8238a780db45}{J. Exp. Theor. Phys. \textbf{20}, 1307 (1965)}.

\bibitem{SLAC1}
D.~L.~Burke, R.~C.~Field, G.~HortonSmith, J.~E.~Spencer, D. Walz, S.~C.~Berridge, W.~M.~Bugg, K.~Shmakov, A.~W.~Weidemann, C.~Bula, et al.,
Positron production in multiphoton light-by-light scattering,
\href{https://journals.aps.org/prl/abstract/10.1103/PhysRevLett.79.1626}
{Phys. Rev. Lett. \textbf{79}, 1626 (1997).}

\bibitem{SLAC2}
C.~Bamber, S.~J.~Boege, T.~Koffas,  T.~Kotseroglou,  A.~C.~Melissinos, D.~D.~Meyerhofer, D.~A.~Reis, W.~Ragg, C.~Bula,  K.~T.~McDonald, et al.,
Studies of nonlinear QED in collisions of 46.6 GeV electrons with intense laser pulses,
\href{https://journals.aps.org/prd/abstract/10.1103/PhysRevD.60.092004}
{Phys. Rev. D \textbf{60}, 092004 (1999).}

\bibitem{Schutzhold:2008pz}
R.~Sch\"utzhold, H.~Gies, and G.~Dunne,
Dynamically assisted Schwinger mechanisms,
\href{https://journals.aps.org/prl/abstract/10.1103/PhysRevLett.101.130404}
{Phys. Rev. Lett. \textbf{101}, 130404 (2008)}.

\bibitem{Taya:2020}
H.~Taya,
Dynamically assisted Schwinger mechanism and chirality production in parallel electromagnetic field,
\href{https://journals.aps.org/prresearch/abstract/10.1103/PhysRevResearch.2.023257}
{Phys. Rev. Res. \textbf{2}, 023257 (2020)}.

\bibitem{CK:2021}
C.~Kohlf\"urst, F,~Queisser, R.~Sch\"utzhold,
Dynamically assisted tunneling in the impulse regime
\href{https://journals.aps.org/prresearch/abstract/10.1103/PhysRevResearch.3.033153}
{Phys. Rev. Res. \textbf{3}, 033153 (2022)}.

\bibitem{Aleksandrov2018}
I.~A.~Aleksandrov, G.~Plunien, and V.~M.~Shabaev,
Dynamically assisted Schwinger effect beyond spatially uniform field approximation,
\href{https://journals.aps.org/prd/abstract/10.1103/PhysRevD.97.116001}
{Phys. Rev. D \textbf{97}, 116001 (2018)}.

\bibitem{IbrahimSitiwaldi2018}
I.~Sitiwaldi and B.~S.~Xie,
The pair production by the three fields dynamically assists the Schwinger process,
\href{https://www.sciencedirect.com/science/article/pii/S0370269317310468}
{Phys. Lett. B \textbf{777}, 406 (2018)}.

\bibitem{Dumlu2010}
C.~K.~Dumlu,
Schwinger Vacuum Pair Production in Chirped Laser Pulses,
\href{https://journals.aps.org/prd/abstract/10.1103/PhysRevD.82.045007}
{Phys. Rev. D \textbf{82}, 045007 (2010)}.

\bibitem{chirped1}
O.~Oluk, B.~S.~Xie, M.~A.~Bake, S.~Dulat,
Electron-positron pair production in strong asymmetric laser electric field,
\href{https://link.springer.com/article/10.1007/s11467-013-0379-8}
{Front. Phys. (Beijing) \textbf{9} 157 (2014).}

\bibitem{chirped2}
O.~Olugh, Z.~L.~Li, B.~S.~Xie, and R.~Alkofer,
Pair production in differently polarized electric fields with frequency chirps,
\href{https://journals.aps.org/prd/abstract/10.1103/PhysRevD.99.036003}
{Phys. Rev. D \textbf{99}, 036003 (2019)}.

\bibitem{chirped3}
O.~Olugh, Z.~L.~Li, and B.~S.~Xie,
Effects of asymmetric pulses on pair production in polarized electric fields,
\href{https://www.cambridge.org/core/journals/high-power-laser-science-and-engineering/
article/asymmetric-pulse-effects-on-pair-production-in-polarized-electric-fields/
0E2AE4AD1DE83B71CD466DAF8A1482FB}
{High Power Laser Sci. Eng. \textbf{8}, e38 (2020)}.

\bibitem{chirped4}
B.~S.~Xie, L.~J.~Li, M.~Mohamedsedik, L.~Wang,
Enhancement effect of frequency chirp on vacuum electron-positron pair production in a strong field,
\href{https://wulixb.iphy.ac.cn/article/doi/10.7498/aps.71.20220148}
{Acta Phys. Sin. \textbf{71}, 131201 (2022)}.

\bibitem{chirped5}
M.~Ababekri, S.~Dulat, B.~S.~Xie and J.~Zhang
Chirp effects on pair production in oscillating electric fields with spatial inhomogeneity,
\href{https://linkinghub.elsevier.com/retrieve/pii/S0370269320306183}
{Phys. Lett. B \textbf{810}, 135815 (2020)}.

\bibitem{chirped6}
C.~Gong, Z.~L.~Li, B.~S.~Xie, and Y.~J.~Li,
Electron-positron pair production in the frequency-modulated laser fields,
\href{https://journals.aps.org/prd/abstract/10.1103/PhysRevD.101.016008}
{Phys. Rev. D \textbf{101}, 016008 (2020)}.

\bibitem{chirped7}
N.~Z.~Chen, O.~Amat, L.~N.~Hu, H.~H.~Fan, and B.~S.~Xie,
Asymmetric pulse effects on pair production in chirped electric fields,
\href{https://journals.aps.org/prd/abstract/10.1103/PhysRevD.109.076015}
{Phys. Rev. D \textbf{109}, 076015 (2024)}.

\bibitem{vortices2}
Z.~L.~Li, Y.~J.~Li, and B.~S.~Xie,
Momentum Vortices in Pair Production using Two Counter-Rotating Fields,
\href{https://journals.aps.org/prd/abstract/10.1103/PhysRevD.96.076010}
{Phys. Rev. D \textbf{96}, 076010 (2017)}.

\bibitem{vortices1}
Z.~L.~Li, D.~Lu, B.~S.~Xie, B.~F.~Shen, L.~B.~Fu, and J.~Liu,
Nonperturbative signatures in pair production for general elliptic polarization fields,
\href{https://iopscience.iop.org/article/10.1209/0295-5075/110/51001}
{Europhys. Lett. 110, 51001 (2015)}.

\bibitem{vortices3}
Z.~L.~Li, B.~S.~Xie, and Y.~J.~Li,
Vortices in multiphoton pair production by two-color rotating laser fields,
\href{https://iopscience.iop.org/article/10.1088/1361-6455/aaf3f9}
{J. Phys. B \textbf{52}, 025601 (2018)}.

\bibitem{vortices4}
L.~N.~Hu, O.~Amat, L.~Wang, A.~Sawut, H.~H.~Fan, and B.~S.~Xie,
Momentum spirals in multiphoton pair production were revisited,
\href{https://journals.aps.org/prd/abstract/10.1103/PhysRevD.107.116010}
{Phys. Rev. D \textbf{107}, 116010 (2023)}.

\bibitem{Strobel2015}
E.~Strobel and S.~S.~Xue,
Semiclassical pair production rate for rotating electric fields,
\href{https://journals.aps.org/prd/abstract/10.1103/PhysRevD.91.045016}
{Phys. Rev. D \textbf{91}, 045016 (2015)}.

\bibitem{Blinne2016}
A.~Blinne and E.~Strobel,
Comparison of semiclassical and Wigner function methods for pair production in rotating fields,
\href{https://journals.aps.org/prd/abstract/10.1103/PhysRevD.93.025014}
{Phys. Rev. D \textbf{93}, 025014 (2016)}.

\bibitem{CKspin2019}
C.~Kohlf\"urst,
Spin states in multiphoton pair production for circularly polarized light,
\href{https://journals.aps.org/prd/abstract/10.1103/PhysRevD.99.096017}
{Phys. Rev. D \textbf{99}, 096017 (2019)}.

\bibitem{hulinaspin}
L.~N.~Hu, H.~H.~Fan, O.~Amat, S.~Tang, and B.~S.~Xie,
Spin-effect-induced momentum spiral and asymmetry degree in pair production,
\href{https://journals.aps.org/prd/abstract/10.1103/PhysRevD.110.056013}
{Phys. Rev. D \textbf{110}, 056013 (2024)}.

\bibitem{spin1}
D.~Seipt and B.~King,
Spin-and polarization-dependent local constant-field approximation rates for nonlinear Compton and Breit-Wheeler processes,
\href{https://journals.aps.org/pra/abstract/10.1103/PhysRevA.102.052805}
{Phys. Rev. A \textbf{102}, 052805 (2020)}.

\bibitem{spin2}
S.~Tang,
Fully polarized nonlinear Breit-Wheeler pair production in pulsed plane waves,
\href{https://journals.aps.org/prd/abstract/10.1103/PhysRevD.105.056018}
{Phys. Rev. D \textbf{105}, 056018 (2022)}.

\bibitem{spin3}
Y.~Y.~Chen, K.~Z.~Hatsagortsyan, C.~H.~Keitel, and R.~Shaisultanov,
Electron spin- and photon-polarization-resolved probabilities of the strong-field QED processes,
\href{https://journals.aps.org/prd/abstract/10.1103/PhysRevD.105.116013}
{Phys. Rev. D \textbf{105}, 116013 (2022)}.

\bibitem{spin4}
J.~J.~Jiang, Y.~N.~Dai, K.~H.~Zhuang, Y.~Gao, S.~Tang, and Y.~Y.~Chen,
Interferences effects in the polarized nonlinear Breit-Wheeler process,
\href{https://journals.aps.org/prd/abstract/10.1103/PhysRevD.109.036030}
{Phys. Rev. D \textbf{109}, 036030 (2024)}.

\bibitem{spin5}
S.~Tang, Y.~Xin, M.~Wen, M.~A.~Bake, and B.~S.~Xie,
Fully polarized Compton scattering in plane waves and its polarization transfer,
\href{https://pubs.aip.org/aip/mre/article/9/3/037204/3279506/Fully-polarized-Compton-scattering-in-plane-waves}
{Matter Radiat. Extremes \textbf{9}, 037204 (2024)}.


\bibitem{spin11}
D.~Y.~Ivanov, G.~L.~Kotkin, V.~G.~Serbo,
 Complete description of polarization effects in e+ e- pair productionby a photon in the field of a strong laser wave,
\href{https://link.springer.com/article/10.1140/epjc/s2005-02125-1#citeas}
{Eur. Phys. J. C \textbf{40}, 27-40 (2005).}

\bibitem{spin12}
T.~M\"uller,  C.~M\"uller,
Longitudinal spin polarization in multiphoton Bethe-Heitler pair production,
\href{https://journals.aps.org/pra/abstract/10.1103/PhysRevA.86.022109}
{Phys. Rev. A \textbf{86}, 022109 (2012).}


\bibitem{spin13}
M. J. A.~Jansen, J. Z.~Kami\ifmmode \acute{n}\else \'{n}\fi{}ski, K.~Krajewska, and C.~M\"uller,
Strong-field Breit-Wheeler pair production in short laser pulses: Relevance of spin effects,
\href{https://link.aps.org/doi/10.1103/PhysRevD.94.013010}
{Phys. Rev. D \textbf{94}, 013010 (2016).}

\bibitem{spin14}
Y.~Y.~Chen, P.~L.~He, R.~Shaisultanov,  K.~Z.~Hatsagortsyan,  C.~H.~Keitel,
Polarized Positron Beams via Intense Two-Color Laser Pulses,
\href{https://journals.aps.org/prl/abstract/10.1103/PhysRevLett.123.174801}
{Phys. Rev. Lett. \textbf{123}, 174801 (2019).}

\bibitem{spin15}
Y.~F.~Li, Y.~Y.~Chen, W.~M.~Wang,  H.~S.~Hu,
Production of Highly Polarized Positron Beams via Helicity Transfer from Polarized Electrons in a Strong Laser Field,
\href{https://journals.aps.org/prl/abstract/10.1103/PhysRevLett.125.044802}
{Phys. Rev. Lett. \textbf{125}, 044802 (2020).}

\bibitem{spin16}
T.~N.~Wistisen,
Numerical approach to the semiclassical method of pair production for arbitrary spins and photon polarization,
\href{https://journals.aps.org/prd/abstract/10.1103/PhysRevD.101.076017}
{Phys. Rev. D \textbf{101}, 076017 (2020).}



\bibitem{spin6}
I.~A.~Aleksandrov and A.~Kudlis,
Pair production in rotating electric fields via quantum kinetic equations:
Resolving helicity states,
\href{https://journals.aps.org/prd/abstract/10.1103/PhysRevD.100.016003}
{Phys. Rev. D \textbf{110}, L011901 (2024)}.

\bibitem{spin7}
A.~W\"ollert, H.~Heiko, and C.~H.~Keitel,
Spin polarized electron-positron pair production via elliptical polarized laser fields,
\href{https://journals.aps.org/prd/abstract/10.1103/PhysRevD.91.125026}
{Phys. Rev. D \textbf{91}, 125026 (2015)}.

\bibitem{spin8}
X.~G.~Huang, M.~Matsuo, and H.~Taya,
Spontaneous generation of spin current from the vacuum by strong electric fields,
\href{https://academic.oup.com/ptep/article/2019/11/113B02/5626098?login=false}
{Prog. Theor. Exp. Phys. \textbf{11}, 113B02 (2019)}.

\bibitem{Review}
A.~Fedotov, A.~Ilderton b, F.~Karbstein, B.~King, D.~Seipt, H.~Taya, G.~ Torgrimsson,
Advances in QED with intense background fields,
\href{https://www.sciencedirect.com/science/article/pii/S0370157323000352?via3Dihub}
{Phys. Rep. \textbf{1010}, 1-138 (2023)}.


\bibitem{RINGWALD2001107}
A. Ringwald,
Pair production from vacuum at the focus of an X-ray free electron laser,
\href{https://www.sciencedirect.com/science/article/pii/S0370269301004968}
{Phys. Lett. B \textbf{510}, 107 (2001).}

\bibitem{DHW1}
F.~Hebenstreit, R.~Alkofer, and H.~Gies,
Particle self-bunching in the Schwinger effect in spacetime-dependent electric fields,
\href{https://journals.aps.org/prl/abstract/10.1103/PhysRevLett.107.180403}
{Phys. Rev. Lett. \textbf{107}, 180403 (2011)}.

\bibitem{DHW3}
M.~Ababekri, B.S.~Xie, J.~Zhang,
Effects of finite spatial extent on Schwinger pair production,
\href{https://journals.aps.org/prd/abstract/10.1103/PhysRevD.100.016003}
{Phys. Rev. D \textbf{100}, 016003 (2019)}.

\bibitem{DHW4}
Z.~L.~Li, C.~Gong, and Y.~J.~Li,
Study of pair production in inhomogeneous two-color electric fields using computational quantum field theory,
\href{https://journals.aps.org/prd/abstract/10.1103/PhysRevD.103.116018}
{Phys. Rev. D \textbf{103}, 116018 (2021)}.

\bibitem{DHW5}
E.~Osman, J.~H.~Bai, Z.~Y.~Chen, M.~A.~Bake,
Efficient vacuum pair production in frequency chirped electric fields with spatiotemporal inhomogeneity,
\href{https://www.sciencedirect.com/science/article/pii/S2211379723009889}
{Results in Phys. \textbf{55}, 107195 (2023)}.

\bibitem{DHW6}
M.~A.~Bake,
Enhanced vacuum pair production by combination of two spatially separated
electric fields,
\href{https://www.sciencedirect.com/science/article/pii/S2211379724001682}
{Results in Phys. \textbf{58}, 1071486 (2024)}.


\bibitem{Blinne2014}
A.~Blinne and H.~Gies,
Pair Production in Rotating Electric Fields,
\href{https://journals.aps.org/prd/abstract/10.1103/PhysRevD.89.085001}
{Phys. Rev. D \textbf{89}, 085001 (2014)}.

\bibitem{1991}
I.~Bialynicki-Birula, P.~Gornicki, J.~Rafelski,
Phase-pace structure of the Dirac vacuum,
\href{https://journals.aps.org/prd/abstract/10.1103/PhysRevD.44.1825}
{Phys. Rev. D \textbf{44}, 1825 (1991).}

\bibitem{CKPHD}
C.~Kohlf\"urst, Electron-positron pair production in inhomogeneous electromagnetic fields, Ph.D. thesis, University of Graz, 2015,
\href{https://arxiv.org/abs/1512.06082}{[arXiv:1512.06082 [hep-ph]]}.

\bibitem{HebenstreitPHD}
F. Hebenstreit, Schwinger effect in inhomogeneous electric fields, Ph.D. thesis, University of Graz, 2011.
\href{https://arxiv.org/abs/1106.5965}{[arXiv:1106.5965 [hep-ph]]}.

\bibitem{Amat:2023vwv}
O.~Amat, L.~N.~Hu, M.~A.~Bake, M.~Mohamedsedik, and B.~S.~Xie,
Effects of spatially oscillating fields on Schwinger pair production,
\href{https://journals.aps.org/prd/abstract/10.1103/PhysRevD.108.056011}
{Phys. Rev. D \textbf{108}, 056011 (2023)}.


\bibitem{Majczak:2024hmt}
M.~M.~Majczak, K.~Krajewska, J.~Z.~Kami\'nski, and A.~Bechler,
Scattering matrix approach to dynamic Sauter-Schwinger process: spin- and helicity-resolved momentum distributions,
\href{https://journals.aps.org/prd/abstract/10.1103/PhysRevD.110.116025}
{Phys. Rev. D \textbf{110}, 116025 (2024)}.

\bibitem{Aleksandrov:2024rsz}
I.~A.~Aleksandrov, A.~Kudlis, and A. I. Klochai,
Kinetic theory of vacuum-pair production in uniform electric fields,
\href{https://journals.aps.org/prresearch/abstract/10.1103/PhysRevResearch.6.043009}
{Phys. Rev. Res. \textbf{6}, 043009 (2024)}.


\bibitem{Amat:2024nvg}
O.~Amat, H.~H.~Fan, S.~Tang, Y.~F.~Huang, and B.~S.~Xie,
Spin-resolved momentum spectra for vacuum pair production using a generalized two-level model,
\href{https://arxiv.org/abs/2409.11833}{[arXiv:2409.11833 [hep-ph]]}.


\bibitem{ck2018}
C.~Kohlf\"urst,
Phase-space analysis of Schwinger effect in inhomogeneous electromagnetic fields,
\href{https://link.springer.com/article/10.1140/epjp/i2018-12062-6}
{Eur. Phys. Journal of Plus \textbf{133}, 191 (2018)}.

\bibitem{Effectivemass}
C.~Kohlf\"urst, H.~Gies, and R.~Alkofer,
Effective mass signatures in multiphoton pair production,
\href{https://journals.aps.org/prl/abstract/10.1103/PhysRevLett.112.050402}
{Phys. Rev. Lett. \textbf{112}, 050402 (2014)}.

\end{thebibliography}
\end{document}